# A New Secret Key Agreement Scheme in a Four-Terminal Network

Parisa Babaheidarian
ISSL Lab., Dept. of Electrical Engineering
Sharif University of Technology
Tehran, Iran
babaheidarian@ee.sharif.edu

Somayeh Salimi
ISSL Lab., Dept. of Electrical Engineering
Sharif University of Technology
Tehran, Iran
salimi@ee.sharif.edu

Mohammad Reza Aref
ISSL Lab., Dept. of Electrical Engineering
Sharif University of Technology
Tehran, Iran
aref@sharif.edu

**Abstract: A new scenario for generating a secret key and two private keys among three Terminals in the presence of an external eavesdropper is considered. Terminals 1, 2 and 3 intend to share a common secret key concealed from the external eavesdropper (Terminal 4) and simultaneously, each of Terminals 1 and 2 intends to share a private key with Terminal 3 while keeping it concealed from each other and from Terminal 4. All four Terminals observe i.i.d. outputs of correlated sources and there is a public channel from Terminal 3 to Terminals 1 and 2. An inner bound of the "secret key-private keys capacity region" is derived and the single letter capacity regions are obtained for some special cases.**

## I. INTRODUCTION[1]

The broadcasting nature of wireless communication networks leads to easy eavesdropping. Due to this fact, secret key sharing between Terminals is of great importance to provide secure communication. An approach to achieve this goal is based on exploiting correlated sources as common randomness between Terminals and communicating over a public channel.

Communication of confidential message was characterized by Wyner [1], in which the model of wiretap channel was introduced. Subsequently, Csiszar and Korner in [2] considered the problem of transmitting a confidential message with a common message in a non-degraded discrete memoryless broadcast channel. Sharing a secret key between two Terminals in the presence of an eavesdropper was first investigated by Maurer in [3] and Ahlswede and Csiszar in [4] where the source and the channel common randomness were used to share a secret key between the two Terminals. Csiszar and Narayan in [5] considered the mentioned problem in the presence of a forth node as a helper. Sharing a secret key between more than two Terminals has been explored by several authors such as [6-14]. In some other scenarios, it is intended to share multiple secret keys among different groups of Terminals simultaneously. In [11] Ye and Narayan examined the problem of sharing a secret-key and a private-key for three Terminals. They investigated the problem for generating a secret key among three Terminals and simultaneously a designated pair of Terminals generates a private key. They explored the problem at the presence of an eavesdropper who can just wiretap the communications over the public channel and has not any other resources to eavesdrop. In [13] Salimi et.al investigated generating two private-keys in a source model consisting of three users. Each of users 1 and 2 intends to share a private key with user 3 where user 1 acts as a wire tapper for user 2 and vice versa. In their scenario, two situations were investigated based on direction of public channel.

In this paper, a new scenario is considered for sharing one secret key and two private keys, simultaneously. In our scheme, there are four Terminals that observe i.i.d. outputs of distinct correlated sources. Terminals 1, 2 and 3 wish to share a secret key among them which should be effectively concealed from the forth Terminal which acts as an external eavesdropper. Simultaneously, each of Terminals 1 and 2 intends to share a private key with Terminal 3 while keeping it concealed from each other and from the external eavesdropper. There is a noiseless public channel of unlimited capacity from Terminal 3 to the other Terminals. Compared to [11], our work requires different security requirements due to the fact that the external wire tapper is equipped with source observations rather than the public channel observations. Also, sharing two private keys are intended in our work. Beside, compared to [12,13], our work necessitates another requirement as sharing a common secret key between the three Terminals and considering an external eavesdropper. We have derived an inner bound of the secret key-private keys capacity region in the described scenario. It is not known if this bound is the capacity region in general, however, we have shown that the inner bound is the capacity region in some special cases which means that our bound is tight.

The organization of this paper is as follows. In Section II, the model and the problem preliminaries are illustrated. In

[1] . This work was partially supported by Iranian NFS under Contract No.88.114/46 and also by Iran telecommunication research center (ITRC).

Section III, our main results and the intuition behind them are given. In Section IV, the proofs are provided. Finally, in Section V, the concluding remarks are given. Throughout the paper, the upper case letters indicate random variables and the lower case letters indicate their realizations.

## II. PROBLEM SETUP

Consider a discrete memoryless multiple sources (DMMS) consisting of four components with alphabets $(\chi_1,\chi_2,\chi_3,\chi_4)$ and corresponding generic random variables (r.v.s) $X_1,X_2,X_3,X_4$. Terminals 1, 2, 3 and 4, respectively, observe n i.i.d. repetitions of the r.v.s $X_1,X_2,X_3,X_4$. Furthermore, a noiseless public channel of unlimited capacity is available such that Terminal 3 can broadcast its required information over it. The information sent over the public channel is depicted by $F_3$ which is a stochastic function of Terminal 3's observation or in other words $F_3=f(X_3^N)$.

The r.v.s $K_0,K_1,K_2$ are all functions of $X_3^N$ taking values from the finite sets $\mathrm{K}_0,\mathrm{K}_1,\mathrm{K}_2$, respectively. $K_0$ represents the common secret key among Terminals 1 and 2 and 3 and $K_1,K_2$ represent private keys for Terminals 1, 2 respectively. Terminal 3 generates a secret key and two private keys $(SK,PK_1,PK_2)$ as a function of $X_3^N$ or in other words $(SK,PK_1,PK_2)=f^*(X_3^N)$ and then, Terminal 3 sends $F_3$ over the public channel. After receiving $F_3$, Terminals 1 and 2 compute the estimation of key pairs $(K_0,K_1),(K_0,K_2)$, respectively, as deterministic functions of the information available at each Terminal, such that:

$$(\hat{K}_{0,1},\hat{K}_1)=g_1(X_1^N,F_3),\qquad (\hat{K}_{0,2},\hat{K}_2)=g_2(X_2^N,F_3)$$

Here $(\hat{K}_{0,1},\hat{K}_{0,2})$ take value from the finite set $\mathrm{K}_0$, and $(\hat{K}_1,\hat{K}_2)$ take value from the finite sets $\mathrm{K}_1,\mathrm{K}_2$, respectively.

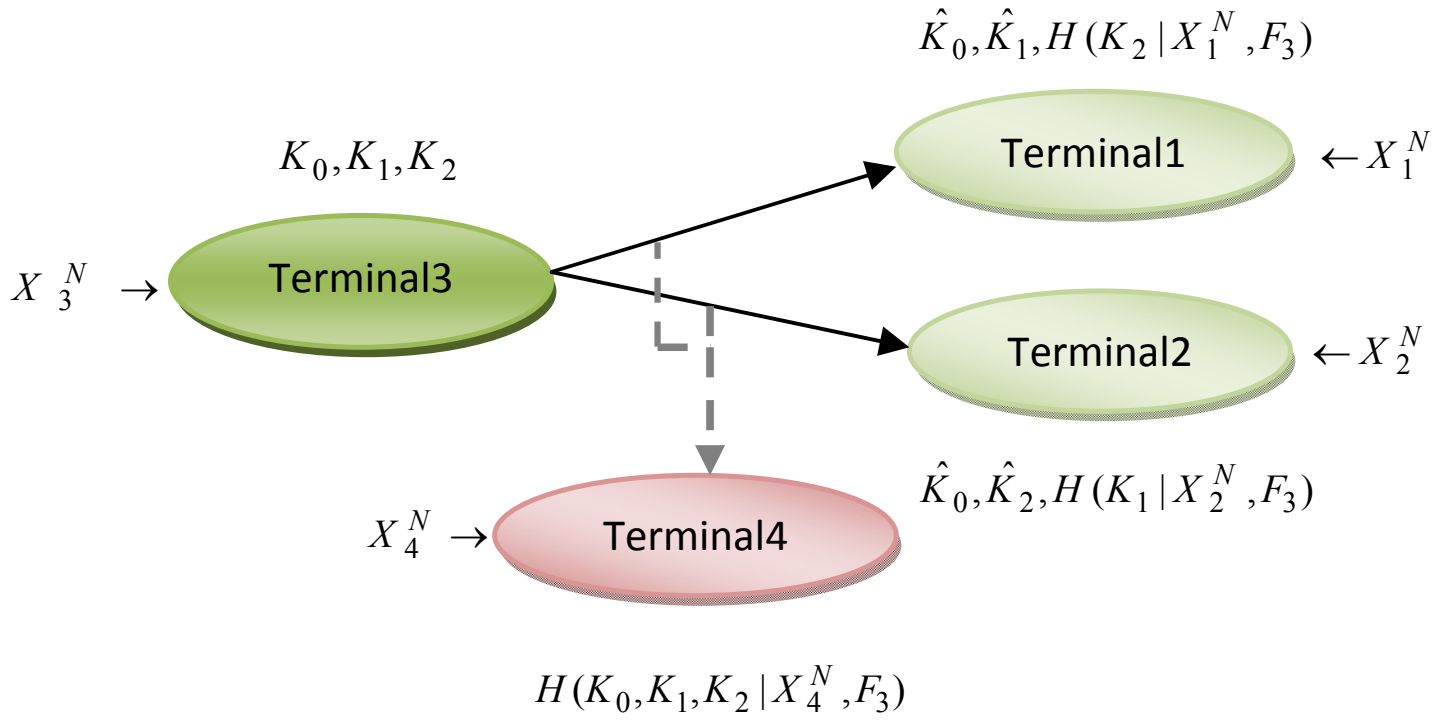

Fig1. Key agreement between Terminals

Now we define conditions which should be addressed in our key sharing scheme.

***Definition 1***: The secret key-private keys rate triple $(R_0,R_1,R_2)$ is an achievable rate triple, if for every $\varepsilon>0$ and $N$ sufficiently large, we have:

$$\Pr\{\bigcup_{i=1}^{2}\{\hat{k}_{0,i}\neq k_0\}\}<\varepsilon \tag{1}$$

$$\Pr\{\hat{k}_1\neq k_1\}<\varepsilon,\Pr\{\hat{k}_2\neq k_2\}<\varepsilon \tag{2}$$

$$\frac{1}{N}I(K_2;X_1^N,F_3)<\varepsilon,\ \frac{1}{N}I(K_1;X_2^N,F_3)<\varepsilon \tag{3}$$

$$\frac{1}{N}I(K_0,K_1,K_2;X_4^N,F_3)<\varepsilon \tag{4}$$

$$\frac{1}{N}\log|K_0|<\frac{1}{N}H(K_0)+\varepsilon \tag{5}$$

$$\frac{1}{N}\log|K_2|<\frac{1}{N}H(K_2)+\varepsilon,\frac{1}{N}\log|K_1|<\frac{1}{N}H(K_1)+\varepsilon, \tag{6}$$

$$\frac{1}{N}H(K_0)>R_0-\varepsilon,\frac{1}{N}H(K_1)>R_1-\varepsilon,\frac{1}{N}H(K_2)>R_2-\varepsilon \tag{7}$$

Equations (1) and (2) are the reliability conditions of the keys at Terminals 2 and 3. Equation (3) means that the private keys of Terminals 2 and 3 are effectively hidden from each other. Equation (4) means that all the keys should be kept secret from Terminal 4. Finally the set of equations in (5-6) indicate the uniformity conditions.

***Definition 2***: The region containing all the achievable secret key-private keys rate triple $(R_0,R_1,R_2)$ is the secret key-private keys capacity region.

## III. STATEMENT OF RESULTS

In this section we state our main results.

***Theorem 1:*** **(*inner bound*)** The following region of non-negative rate triples is achievable for the described source model:

$$R^I=Conv\bigcup_{P(u_0,u_1,u_2,x_1,x_2,x_3,x_4,q)}$$

$$\left\{\begin{array}{l}(R_0,R_1,R_2):\\ R_0\geq 0,R_1\geq 0,R_2\geq 0\\ R_0\leq[\min\{I(U_0;X_1|Q),I(U_0;X_2|Q)\}-I(U_0;X_4|Q)]^+\\ R_1\leq[I(U_1;X_1|U_0,Q)-\\ \max\{I(U_1;X_2,U_2|U_0,Q),I(U_1;X_4,U_2|U_0,Q)\}]^+\\ R_2\leq[I(U_2;X_2|U_0,Q)-\\ \max\{I(U_2;X_1,U_1|U_0,Q),I(U_2;X_4,U_1|U_0,Q)\}]^+\end{array}\right\} \tag{8}$$

where $Q,U_0,U_1,U_2$ are random variables taking values in sufficiently large finite sets and according to the distribution:

$$p(u_0,u_1,u_2,x_1,x_2,x_3,x_4,q)=p(q|u_0,u_1,u_2)p(u_0|x_3)p(u_1|u_0,x_3)p(u_2|u_0,x_3)$$
$$p(x_1,x_2,x_3,x_4)$$

The function $[x]^+$ equals $x$ if $x\geq 0$ and $0$ if $x<0$, and Conv operator is a convex closure of the set. The sketch of the proof is given in Section IV.A. However, in continue, we briefly explain our coding scenario.

***Remark 1***: The above region can be achieved via separate decoding. For achievability, we use a scheme which utilizes superposition coding as well as double layer binning. A similar method, called the secret superposition scheme, was used in [16] to send a confidential message in the channel model. Here, we use superposition coding to handle the secret key. In addition, double layer Binnig is used to meet the secrecy constrains in all layers which means that all the keys should be kept secret from the forth node. Our coding scheme is illustrated in Fig. 2.

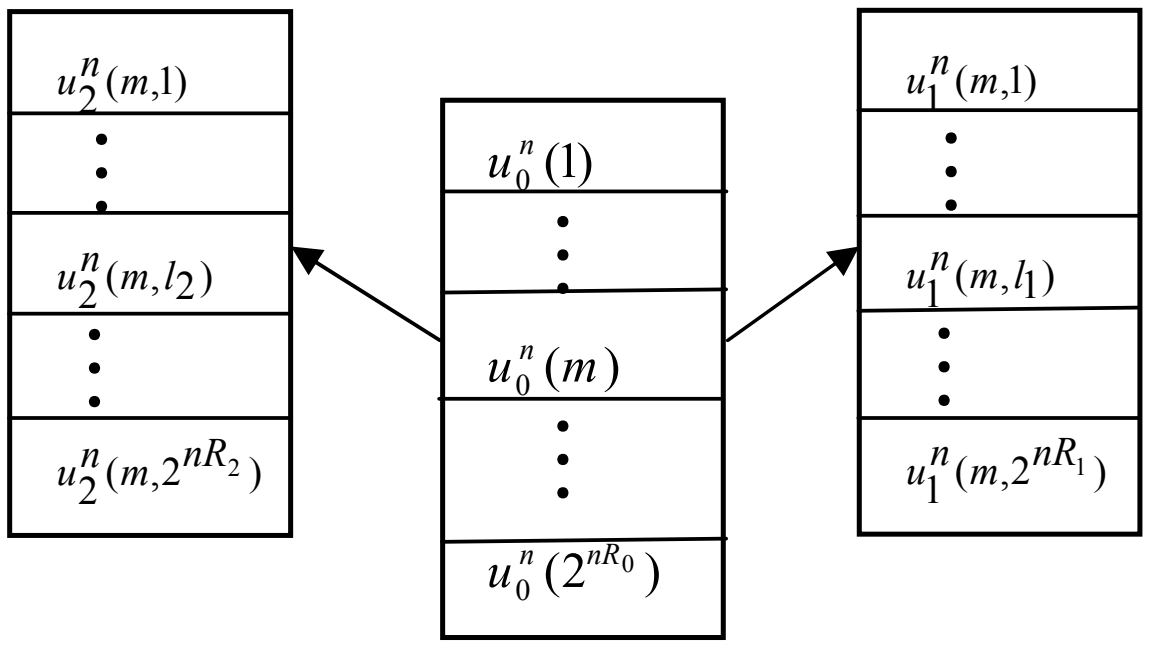

Fig. 2. Coding scheme

***Remark 2***: The dual problem of the described model in channel model is a broadcast channel with three receivers, in which different levels of secrecy must be satisfied and this problem is an open problem in general.

We could not prove that the rate region in (8) is the capacity region in general; however, we have investigated some special cases where the rate region in (8) is the secret key-private keys capacity region. First, we derive an explicit outer bound for the secret key-private keys capacity region, and then we provide some special cases where the region in (8) coincides with the explicit outer bound.

***Proposition 1***: In the key agreement scenario of the described model, if the rate triple $(R_0, R_1, R_2)$ is achievable, then it satisfy:

$$\begin{Bmatrix} R_0 \le \min\{I(X_3;X_1|X_4), I(X_3;X_2|X_4)\} \\ R_1 \le \min\{I(X_3;X_1|X_4), I(X_3;X_1|X_2)\} \\ R_2 \le \min\{I(X_3;X_2|X_4), I(X_3;X_2|X_1)\} \end{Bmatrix} \qquad (9)$$

for any joint distribution $P(X_1, X_2, X_3, X_4)$. This bound can be automatically deduced from Theorem 1 of [4].

***Corollary 1***: When the source observations form a Markov chain as $X_3 - X_1 - X_4 - X_2$, the secret key-private keys capacity region reduces to:

$$R_0 = 0, R_2 = 0, 0 \le R_1 \le I(X_3;X_1|X_4).$$

Achievability follows from Theorem 1 by setting $U_1 = X_3, U_0 = U_2 = Q = \varnothing$ and noting that $R_1 = I(X_3;X_1) - I(X_3;X_4) = I(X_3;X_1|X_4)$. The converse can be deduced from Proposition 1. It should be noted that when the source observations form a Markov chain as $X_3 - X_2 - X_4 - X_1$, the Secret key-private keys capacity region can be derived by symmetry from Corollary 1.

***Corollary 2***: When the source observations form a Markov chain as $X_1 - X_3 - X_4 - X_2$, the secret key-private keys capacity region reduces to:

$$R_0 = 0, R_2 = 0, 0 \le R_1 \le I(X_3;X_1|X_4).$$

Achievability proof is the same as Corollary 1. The converse proof directly follows from the explicit upper bound of Proposition 1. When the source observations form a Markov chain as $X_2 - X_3 - X_4 - X_1$, the secret key-private keys capacity region can be derived by symmetry from Corollary 2.

***Corollary 3***: When the source observations form a Markov chain as $X_3 - X_1 - X_2 - X_4$, the secret key-private keys capacity region reduces to:

$$0 \le R_0 \le I(U_0;X_2|Q) - I(U_0;X_4|Q),$$
$$0 \le R_1 \le I(U_1;X_1|U_0,Q) - I(U_1;X_2|U_0,Q), R_2 = 0$$

where $U_1, U_0, Q$ are auxiliary random variables taking values in sufficiently large finite sets and according to the distribution $p(u_1,u_0,q,x_1,x_2,x_3,x_4) = p(q|u_1,u_0)p(u_1,u_0|x_3)p(x_1,x_2,x_3,x_4)$ that form Markov chains as $Q - U_0 - X_3, Q - U_1 - X_3$.

Achievability follows from Theorem 1 and the Markov chain $X_3 - X_1 - X_2 - X_4$. The converse is proved in section IV.B. When the source observations form a Markov chain as $X_3 - X_2 - X_1 - X_4$, the Secret key-private keys capacity region can be derived by symmetry from Corollary 3.

***Corollary 4***: When the source observations form a Markov chain as $X_2 - X_1 - X_3 - X_4$, the secret key-private keys capacity region reduces to:

$$0 \le R_0 \le I(U_0;X_2|Q) - I(U_0;X_4|Q),$$
$$0 \le R_1 \le I(U_1;X_1|U_0,Q) - I(U_1;X_4|U_0,Q), R_2 = 0$$

where $U_1, U_0, Q$ are auxiliary random variables taking values in sufficiently large finite sets and according to the distribution $p(u_1,u_0,q,x_1,x_2,x_3,x_4) = p(q|u_1,u_0)p(u_1,u_0|x_3)p(x_1,x_2,x_3,x_4)$ that form Markov chains as $Q - U_0 - X_3, Q - U_1 - X_3$ and $U_1 - (X_4, Q, U_0) - (Q, U_0, X_2)$. Existence of such variables can be deduced from the Markov chain $X_2 - X_1 - X_3 - X_4$. Achievability directly follows from Theorem 1 by setting $U_2 = \varnothing$. The converse can be proved the same as Corollary 3. When the source observations form a Markov chain as $X_1 - X_2 - X_3 - X_4$, the secret key-private keys capacity region can be derived by symmetry from Corollary 4.

***Corollary 5***: When the source observations form a Markov chain as $X_2 - X_3 - X_1 - X_4$, the secret key-private keys capacity region is achievable if :

$0 \le R_0 \le I(U_0;X_2|Q) - I(U_0;X_4|Q),$

$0 \le R_1 \le I(U_1;X_1|U_0,Q) - I(U_1;X_2|U_0,Q),$

$0 \le R_2 \le I(U_2;X_2|U_0,Q) - I(U_2;X_1|U_0,Q)$

where $U_2,U_1,U_0,Q$ are auxiliary random variables taking values in sufficiently large finite sets and according to the distribution $p(u_2,u_1,u_0,q,x_1,x_2,x_3,x_4) = p(q|u_2,u_1,u_0)p(u_2,u_1,u_0|x_3)p(x_1,x_2,x_3,x_4)$ that form Markov chains as $Q-U_0-X_3, Q-U_1-X_3, Q-U_2-X_3, U_2-(U_0,X_3)-U_1$. The proof is similar to the converse proof of corollary 3. This bound coincides with the inner bound if we have :

$$U_0-(X_1,Q)-(X_2,Q) \quad U_1-(U_0,Q,X_1)-(U_0,Q,X_2)-U_2 \quad U_2-(U_1,U_0,Q,X_2)-(U_1,U_0,Q,X_4)$$

This situation is illustrated in Fig.3.

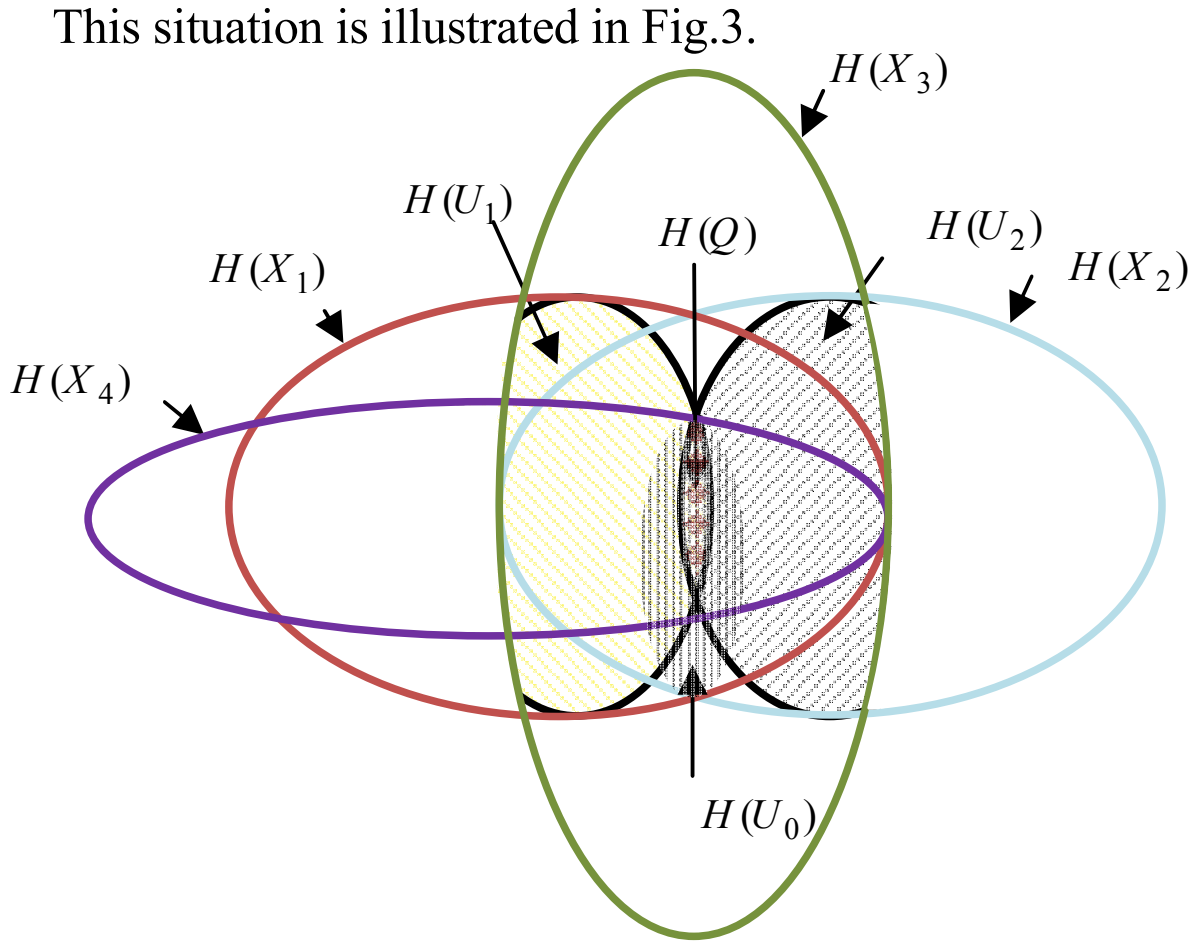

Fig.3. An example for case $X_2 - X_3 - X_1 - X_4$

When the source observations form a Markov chain as $X_1 - X_3 - X_2 - X_4$, the secret key-private keys capacity region can be derived by symmetry from Corollary 5.

For two remaining Markov chains $X_3 - X_4 - X_1 - X_2$ and $X_3 - X_4 - X_2 - X_1$ all the rates will be equal to zero. It can be shown by the explicit upper bound in (9).

## IV. Proofs

### *A. proof of Theorem 1(achieavblity)*

Due to space limitation, we avoid the details of the proof and the proof scheme is given in this section. For more details, it can be referred to the extended version of this paper in [17]. In order to construct the codebook, Terminal 3 generates its random codebook as follow. For an arbitrary distribution $p(u_0)$ generates collection of codewords each uniformly drawn from the set $T_{\varepsilon_1}^{(N)}(p_{U_0|X_3})$. Terminal 3 partitions these codewords into $2^{N(R_0+R_0'+R_0'')}$ bins with the same size in a uniformly manner , where $R_0 + R_0' = H(U_0|X_4,Q) + 2\varepsilon_1$ and $R_0'' = I(U_0;X_4|Q) - \varepsilon_1$. The bin index of each bin is denoted by the triple $(k_0,k_0',k_0'')$ and the corresponding random variables are $K_0,K_0',K_0''$, respectively. Next, for every codeword $u^N_{0,\{k_0,k_0',k_0''\}}$, Terminal 3 generates $2^{N(R_1+R_1'+R_1'')}$ codewords $u_1^N(k_1,k_1',k_1'')$ $k_1 \in \{1,\ldots,2^{NR_1}\}, k_1' \in \{1,\ldots,2^{NR_1'}\}, k_1'' \in \{1,\ldots,2^{NR_1''}\}$, where $R_1 + R_1' = \min\{H(U_1|X_2,U_2,U_0,Q), H(U_1|X_4,U_2,U_0,Q)\} + 2\varepsilon_1$ and $R_1'' = \max\{I(U_1;X_2,U_2|U_0,Q), I(U_1;X_4,U_2|U_0,Q)\} - \varepsilon_1$, according to $\prod_{i=1}^{N} p_{U_1|U_0,X_3}(u_{1,i}|u_{0,i}(k_0,k_0',k_0''),x_{3,i})$ and the corresponding random variables of bin indices are $K_1,K_1',K_1''$. Similarly Terminal 3 generates $2^{N(R_2+R_2'+R_2'')}$ codewords $u_2^N(k_2,k_2',k_2'')$, $k_2 \in \{1,\ldots,2^{NR_2}\}, k_2' \in \{1,\ldots,2^{NR_2'}\}, k_2'' \in \{1,\ldots,2^{NR_2''}\}$, where $R_2 + R_2' = \min\{H(U_2|X_1,U_1,U_0,Q), H(U_2|X_4,U_1,U_0,Q)\} + 2\varepsilon_1$ and $R_2'' = \max\{I(U_2;X_1,U_1|U_0,Q), I(U_2;X_4,U_1|U_0,Q)\} - \varepsilon_1$, according to $\prod_{i=1}^{N} p_{U_2|U_0,X_3}(u_{2,i}|u_{0,i}(k_0,k_0',k_0''),x_{3,i})$ and the corresponding random variables of bin indices are $K_2,K_2',K_2''$. Now, for a fixed distribution $P(q|u_0,u_1,u_2)$ Terminal 3 generates $2^{N(I(U_0U_1U_2;Q))}$ i.i.d. code words of length $N$ with the distribution $p(q)$. All Terminals are informed of binning scheme and distributions. For every ε-typical sequence $X_3^N = x_3^N$, the set of codewords $u_0^N$ which are jointly typical with $x_3^N$ is denoted by $(U_0^N)_{x_3^N}$ and for a fixed jointly typical sequences $(u_0^N,x_3^N)$, the set of codewords $u_1^N$ which are jointly typical with $(u_0^N,x_3^N)$ is denoted by $(U_1^N)_{u_0^N,x_3^N}$. Similarly we can establish the set of codewords $(U_2^N)_{u_0^N,x_3^N}$.

For encoding, Terminal 3 observes the i.i.d. sequence of $X_3^N = x_3^N$ from its memoryless source and selects the corresponding sub-codebook. If the observed sequence was not ε-typical, it would declare an error and tries for the next observation. Afterward, Terminal 3 randomly selects the sequence $u_0^N$ from the set $(U_0^N)_{x_3^N}$ and then randomly selects sequences $(u_1^N,u_2^N)$ from the set $(U_1^N)_{u_0^N,x_3^N},(U_2^N)_{u_0^N,x_3^N}$, respectively. Then He chooses respective row indices, $k_0,k_1,k_2$ of the selected codewords $(u_0^N,u_1^N,u_2^N)$ as a secret key, private key 1 and private key 2 respectively, and he sends the column indices $k_0',k_1',k_2'$ of the codewords over the public channel. In addition Terminal 3 sends index i of $Q^N(i)$ which is jointly typical with the selected sequences $(u_0^N,u_1^N,u_2^N)$.

For decoding, Terminals 1 and 2 should be able to reconstruct random variables pairs $(U_0,U_1)$ and $(U_0,U_2)$ respectively. For the purpose of reliable decoding of sequence $u_0^n$ by Terminals 1 and 2, Terminal 3 should at least send information with the rate $R_0' = \max\{H(U_0|X_1,Q), H(U_0|X_2,Q)\}$, therefore the maximum achievable secret key rate is obtained as:

$$H(U_0) - I(U_0;X_4,Q) - R_0' = \min\{I(U_0;X_1|Q), I(U_0;X_2|Q)\} - I(U_0;X_4|Q),$$

where the term $I(U_0;X_4,Q)+R_0'$ is the leakage information rate. With access to $U_0$, our equivocation from $U_1$ reduces to $H(U_1|U_0)$. From slepian-wolf coding [15], we deduce that Terminal 3 should send information at least with the rate $R_1' = H(U_1|X_1,U_0,Q)$, so that Terminal 1 can correctly reconstruct $U_1$. Hence:

$$R_1 = H(U_1|U_0) - [I(U_1;Q|U_0) + \max\{I(U_1;X_2,U_2|U_0,Q), I(U_1;X_4,U_2|U_0,Q)\} + R_1'].$$

The same approach can be followed for $R_2$. In continue, we will prove security conditions (3). Security condition (4) can be deduced using the same approaches. We have:

$$\begin{aligned}
& I(K_1;X_2^N,F_3) = I(K_1;X_2^N,K_0',K_1',K_2',I) \\
& \le I(K_1;X_2^N,U_0^N,U_2^N,Q^N,K_1') \\
& = H(K_1) - H(U_1^N,K_1|X_2^N,U_0^N,U_2^N,Q^N,K_1') + H(U_1^N|X_2^N,U_0^N,U_2^N,Q^N,K_1',K_1) \\
& \overset{(a)}{=} H(K_1) - H(U_1^N|X_2^N,U_0^N,U_2^N,Q^N,K_1') + H(U_1^N|X_2^N,U_0^N,U_2^N,Q^N,K_1',K_1) \\
& = H(K_1) - H(U_1^N|X_2^N,U_0^N,U_2^N,Q^N) + H(K_1'|X_2^N,U_0^N,U_2^N,Q^N) \\
& + H(U_1^N|X_2^N,U_0^N,U_2^N,Q^N,K_1',K_1) \\
& \le H(K_1) + H(K_1') - H(U_1^N|X_2^N,U_0^N,U_2^N,Q^N) + H(U_1^N|X_2^N,U_0^N,U_2^N,Q^N,K_1',K_1) \\
& = NH(U_1|X_2,U_0,U_2,Q) + 2N\varepsilon_1 - H(U_1^N|X_2^N,U_0^N,U_2^N,Q^N) \\
& + H(U_1^N|X_2^N,U_0^N,U_2^N,Q^N,K_1',K_1) \\
& \overset{(b)}{\le} NH(U_1|X_2,U_0,U_2,Q) + 2N\varepsilon_1 - NH(U_1|X_2,U_0,U_2,Q) + N\varepsilon_2 \\
& + H(U_1^N|X_2^N,U_0^N,U_2^N,Q^N,K_1',K_1) \\
& = 2N\varepsilon_1 + N\varepsilon_2 + H(U_1^N|X_2^N,U_0^N,U_2^N,Q^N,K_1',K_1) \\
& \overset{(c)}{\le} 2N\varepsilon_1 + N\varepsilon_2 + N\varepsilon_3.
\end{aligned}$$

Inequality (a) follows from the fact that $H(K_1|U_1^N,X_2^N,U_0^N,U_2^N,Q^N,K_1') = 0$ because $K_1$ is one of the indices of $U_1^N$, (b) from the same approach as lemma 1 in [14] to show $NH(U_1|X_2,U_0,U_2,Q) \le H(U_1^N|X_2^N,U_0^N,U_2^N,Q^N) + N\varepsilon_2$ and (c) from the same approach as lemma 2 in [14] to show $H(U_1^N|X_2^N,U_0^N,U_2^N,Q^N,K_1',K_1) \le N\varepsilon_3$. By choosing $\varepsilon_1,\varepsilon_2,\varepsilon_3$ such that $\varepsilon_1+\varepsilon_2+\varepsilon_3 \le \varepsilon$, the security condition (3) is proved. Using the same approaches, we can deduce $I(K_2;X_1^N,K_0',K_1',K_2',I) \le \varepsilon$ and hence, the security condition (3) is satisfied. Analysis of security condition in (4) is given in [17].

*B. Converse proof*

In this subsection, the converse part of corollary 3 is proved in which the source observations form the Markov chain $X_3 - X_1 - X_2 - X_4$.

First, we bound the secret key rate $R_0$. Fano's inequality at Terminal 2 results in:

$$H(K_0|X_2^n,F_3) \le 1 + nR_0P_e^{(n)} = n\varepsilon_n$$

Also the security condition at the forth Terminal should be satisfied as $I(K_0,K_1,K_2;X_4^n,F_3) \le n\varepsilon$. We have:

$$\begin{aligned}
nR_0 & \overset{(a)}{\le} H(K_0|X_4^n,F_3) + n\varepsilon \\
& \overset{(b)}{\le} H(K_0|X_4^n,F_3) + n\varepsilon - H(K_0|X_2^n,F_3) + n\varepsilon_n \\
& = I(K_0;X_2^n|F_3) - I(K_0;X_4^n|F_3) + n(\varepsilon+\varepsilon_n) \\
& \overset{(c)}{=} \sum_{i=1}^{n} I(K_0;X_{2,i}|F_3,\tilde{X}_2^{i+1},X_4^{i-1}) - I(K_0;X_{4,i}|F_3,\tilde{X}_2^{i+1},X_4^{i-1}) + n(\varepsilon+\varepsilon_n) \\
& = \sum_{i=1}^{n} I(K_0;X_{2,i}|F_3,\tilde{X}_2^{i+1},X_4^{i-1},\tilde{X}_4^{i+1}) - I(K_0;X_{4,i}|F_3,\tilde{X}_2^{i+1},X_4^{i-1},\tilde{X}_4^{i+1}) \\
& + \sum_{i=1}^{n} I(K_0;\tilde{X}_4^{i+1}|F_3,\tilde{X}_2^{i+1},X_4^{i-1},X_{4,i}) - I(K_0;\tilde{X}_4^{i+1}|F_3,\tilde{X}_2^{i+1},X_4^{i-1},X_{2,i}) \\
& + n(\varepsilon+\varepsilon_n) \\
& \overset{(d)}{\le} \sum_{i=1}^{n} I(K_0;X_{2,i}|F_3,\tilde{X}_2^{i+1},X_4^{i-1},\tilde{X}_4^{i+1}) - I(K_0;X_{4,i}|F_3,\tilde{X}_2^{i+1},X_4^{i-1},\tilde{X}_4^{i+1}) \\
& + n(\varepsilon+\varepsilon_n) \\
& \le \sum_{i=1}^{n} I(K_0;X_{2,i}|X_{4,i},F_3,\tilde{X}_2^{i+1},X_4^{i-1},\tilde{X}_4^{i+1}) + n(\varepsilon+\varepsilon_n) \\
& \le \sum_{i=1}^{n} I(K_0,X_1^{i-1};X_{2,i}|X_{4,i},F_3,\tilde{X}_2^{i+1},X_4^{i-1},\tilde{X}_4^{i+1}) + n(\varepsilon+\varepsilon_n) \\
& \overset{(e)}{=} \sum_{i=1}^{n} I(U_{0,i};X_{2,i}|Q_i,X_{4,i}) + n(\varepsilon+\varepsilon_n) \\
& \overset{(f)}{=} \sum_{i=1}^{n} I(U_{0,i};X_{2,i}|Q_i) - I(U_{0,i};X_{4,i}|Q_i) + n(\varepsilon+\varepsilon_n) \\
& \overset{(g)}{=} n(I(U_{0V};X_{2V}|Q_V) - I(U_{0V};X_{4V}|Q_V) + \varepsilon'),
\end{aligned}$$

where in the above equations, (a) results from the security condition, (b) from Fano's inequality, (c) from Lemma 1 of [2], (d) from the fact that $I(K_0;\tilde{X}_4^{i+1}|F_1,\tilde{X}_3^{i+1},X_4^{i-1},X_{4,i}) = 0$ which can be deduced from the Markov chain $X_3 - X_1 - X_2 - X_4$, and memoryless property, it is shown in lemma 1 in [17], (e) from the definition of the random variables $Q_i, U_{0,i}$ as:

$$Q_i = (F_3,\tilde{X}_2^{i+1},\tilde{X}_4^{i+1},X_4^{i-1}), \qquad U_{0,i} = (K_0,X_1^{i-1},Q_i)$$

(f) from the fact that $I(U_{0,i};X_{4,i}|Q_i,X_{2,i}) = 0$ which can be deduced from the Markov chain $X_3 - X_1 - X_2 - X_4$ and (g) from considering V as a time sharing random variable independent of all the other random variables which uniformly takes value from the set $\{1,2,...,n\}$ and by setting $\varepsilon' \triangleq \varepsilon + \varepsilon_n$.

Now, we will bound $R_1$. According to Fano's inequality for Terminal 1's private key, we have:

$$H(K_1|X_1^n,F_3) \le n\varepsilon_n$$

Also, the security condition for the mentioned private key establishes as:

$$I(K_1;X_2^n,F_3) \le n\varepsilon$$

and consequently:

$$I(K_1;X_2^n,F_3,K_0)=I(K_1;X_2^n,F_3)+I(K_1;K_0\,|\,X_2^n,F_3)$$
$$\le n\varepsilon+H(K_0\,|\,X_2^n,F_3)\le n(\varepsilon+\varepsilon_n)$$

By setting $\varepsilon'\triangleq\varepsilon+\varepsilon_n$ we can conclude $I(K_1;X_2^n,F_3,K_0)\le n\varepsilon'$.

We have:

$$nR_1\overset{(a)}{\le}H(K_1|X_2^n,F_3,K_0)+n\varepsilon'$$
$$\overset{(b)}{\le}H(K_1|X_2^n,F_3,K_0)-H(K_1|X_1^n,F_3)+n\varepsilon'+n\varepsilon$$
$$\le H(K_1|X_2^n,F_3,K_0)-H(K_1|X_1^n,F_3,K_0)+n\varepsilon'+n\varepsilon$$
$$=I(K_1;X_1^n\,|\,F_3,K_0)-I(K_1;X_2^n\,|\,F_3,K_0)+n(2\varepsilon+\varepsilon_n)$$
$$\overset{(c)}{=}\sum_{i=1}^{n}I(K_1;X_{1,i}\,|\,F_3,K_0,X_1^{i-1},\tilde{X}_2^{i+1})-I(K_1;X_{2,i}\,|\,F_3,K_0,X_1^{i-1},\tilde{X}_2^{i+1})+n(2\varepsilon+\varepsilon_n)$$
$$=\sum_{i=1}^{n}I(K_1;X_{1,i}\,|\,F_3,K_0,X_1^{i-1},\tilde{X}_2^{i+1},X_4^{i-1},\tilde{X}_4^{i+1})-I(K_1;X_{2,i}\,|\,F_3,K_0,X_1^{i-1},\tilde{X}_2^{i+1},X_4^{i-1},\tilde{X}_4^{i+1})$$
$$+\sum_{i=1}^{n}I(K_1;X_4^{i-1},\tilde{X}_4^{i+1}\,|\,F_3,K_0,X_1^{i-1},\tilde{X}_2^{i+1},X_{2,i})-I(K_1;X_4^{i-1},\tilde{X}_4^{i+1}\,|\,F_3,K_0,X_1^{i-1},\tilde{X}_2^{i+1},X_{1,i})$$
$$+n(2\varepsilon+\varepsilon_n)$$
$$\overset{(d)}{\le}\sum_{i=1}^{n}I(K_1;X_{1,i}\,|\,F_3,K_0,X_1^{i-1},\tilde{X}_2^{i+1},X_4^{i-1},\tilde{X}_4^{i+1})-I(K_1;X_{2,i}\,|\,F_3,K_0,X_1^{i-1},\tilde{X}_2^{i+1},X_4^{i-1},\tilde{X}_4^{i+1})$$
$$+n(2\varepsilon+\varepsilon_n)$$
$$\overset{(e)}{=}\sum_{i=1}^{n}I(U_{1,i};X_{1,i}\,|\,U_{0,i},Q_i)-I(U_{1,i};X_{2,i}\,|\,U_{0,i},Q_i)+n(2\varepsilon+\varepsilon_n)$$
$$\overset{(f)}{\le}n(I(U_{1V};X_{1V}\,|\,U_{0V},Q_V)-I(U_{1V};X_{2V}\,|\,U_{0V},Q_V))+\varepsilon''$$

where in the above equations, (a) results from the security condition, (b) from Fano's inequality, (c) from lemma 1 in [2], (d) from the fact that $I(K_1;X_4^{i-1},\tilde{X}_4^{i+1}\,|\,F_3,K_0,X_1^{i-1},\tilde{X}_2^{i+1},X_{2,i})=0$ which can be deduced from the Markov chain $X_3-X_1-X_2-X_4$, and memoryless property, it is shown in [17], (e) from the definition of the random variables $U_{1,i}=(K_1,X_1^{i-1},Q_i)$ and $Q_i,U_{0,i}$ as before, and (f) from considering V as a time sharing random variable independent of all the other random variables which uniformly takes value from the set $\{1,2,...,n\}$ and by setting $\varepsilon''\triangleq 2\varepsilon+\varepsilon_n$.

For rate $R_2$, it is clear from the Markov chain that every message Terminal 2 decodes can also be decode by Terminal 1. Therefore in this case the private key 2 rate is equal to zero. But still we can prove this claim by noting that in the explicit outer bound we have $I(X_3;X_2\,|\,X_1)=0$ due to the Markov chain $X_3-X_1-X_2-X_4$.

## V. CONCLUSION

In this paper, we have studied the problem of simultaneous sharing a secret key and two private keys among three Terminals with an additional wire tapper. We have obtained an inner bound to the secret key-private keys capacity region. Also for some special cases, we have shown that the inner bound is the secret key-private keys capacity region demonstrating that the inner bound is tight. The investigated special cases include all the different Markov chains that the sources from.

As a generalization of this work, for the case in which Terminal 3 is not a trusted party, one can study the case of having three private keys in this model. For now, we are working on a case in which the direction of the public channel is reversed; meaning the case in which only Terminals 1 and 2 are allowed to talk over the public channel.